\documentclass[aps,prc,preprint,superscriptaddress,showpacs,floatfix,nobibnotes]{revtex4}
\usepackage{mathrsfs}

\usepackage{bm}
\usepackage{graphicx}
\usepackage{longtable}

\begin{document}

\title{Mass of \emph{Y}(3940) in Bethe-Salpeter equation for quarks }

\author{Xiaozhao Chen}%\email{chen.xzhao.hn@gmail.com} \email[corresponding author]{}
\affiliation{Department of Foundational courses, Shandong University
of Science and Technology, Taian, 271019, China}

\author{Xiaofu L\"{u}}
\affiliation{Department of Physics, Sichuan University, Chengdu,
610064, China} \affiliation{Institute of Theoretical Physics, The
Chinese Academy of Sciences, Beijing 100080, China}
\affiliation{CCAST (World Laboratory), P.O. Box 8730, Beijing
100080, China}

\date{\today}

\begin{abstract}
The general form of the Bethe-Salpeter wave functions for the bound
states composed of two vector fields of arbitrary spin and definite
parity is corrected. Using the revised general formalism, we
investigate the observed \emph{Y}(3940) state which is considered as
a molecule state consisting of $D^{*0}\bar{D}^{*0}$. Though the
attractive potential between $D^{*0}$ and $\bar{D}^{*0}$ including
one light meson ($\sigma$, $\pi$, $\omega$, $\rho$) exchange is
considered, we find that in our approach the contribution from
one-$\pi$ exchange is equal to zero and consider SU(3) symmetry
breaking. The obtained mass of \emph{Y}(3940) is consistent with the
experimental value.
\end{abstract}

\pacs{12.40.Yx, 14.40.Rt, 12.39.Ki}

%\keywords{molecule; Bethe-Salpeter amplitudes; form factor } \maketitle

%\bigskip
\maketitle

\newpage

\parindent=20pt

\section{Introduction}
The exotic state \emph{Y}(3940) was discovered by the Belle
collaboration \cite{SK.CHOI} and then confirmed by the BABAR
collaboration \cite{B.Aubert}. The investigation of the structure of
\emph{Y} is of great significance, while the conventional $c\bar c$
charmonium interpretation for this state is disfavored
\cite{T.Aaltonen}. Then possible alternative interpretations have
been proposed, such as hadronic molecule and tetraquark states.
Following the experimental results, it is suggested in Ref.
\cite{liu} that the \emph{Y}(3940) is a hadronic molecule state of
$D^{*0}\bar{D}^{*0}$. However, in previous work \cite{liu},  the
numerical result of the binding energy for the molecule state
sensitively depends on the value of typical cutoff in the effective
interaction potential between two heavy vector mesons, and this two
heavy mesons are considered as pointlike objects. Furthermore, the
spin-parity quantum numbers $J^P$ of the \emph{Y}(3940) are not
unambiguously determined in experiment, except for $C=+$. The
molecule state hypothesis implies that the quantum numbers of
\emph{Y}(3940) are $0^+$ or $2^+$, but Ref. \cite{liu} can not
deduce the definite quantum numbers in theory.

Though the general form of the Bethe-Salpeter (BS) wave functions
for the bound states consisting of two vector fields of arbitrary
spin and definite parity has been given in Ref. \cite{mypaper3}, we
find that the derivation of this formalism has a serious defect. In
this work, the general formalism is firstly corrected. Then we
assume that the \emph{Y}(3940) state is a molecule state composed of
two heavy vector mesons $D^{*0}\bar{D}^{*0}$ and the revised general
formalism is applied to investigate this two-body system. To
construct the interaction kernel between two heavy mesons derived
from one light meson ($\sigma$, $\pi$, $\omega$, $\rho$) exchange,
we consider that the heavy meson is not a point-like particle but a
bound state composed of u-quark and c-quark and then investigate the
light meson interaction with the u-quark in the heavy meson. Through
the form factor we can obtain the light meson interaction with heavy
meson and the potential between two heavy mesons without an extra
parameter. \cite{mypaper,mypaper2} Obviously, this potential in our
approach contains more inspirations of quantum chromodynamics (QCD).
Finally, numerically solving the relativistic
Schr$\ddot{o}$dinger-like equation with this potential, we can
obtain the mass of the molecule state and then deduce the definite
quantum numbers of the \emph{Y}(3940) system.

In this work, one-$\pi$ exchange is considered in the interaction
kernel between two heavy mesons. When investigating this
pseudoscalar meson interaction with the u-quark in the heavy meson,
we find that the coupling $\mathscr{L}_I=ig_\pi\bar{u}\gamma_5u\pi$
should have no contribution to this interaction and represent it as
the derivative coupling Lagrangian
$\mathscr{L}_I=if_\pi\bar{u}\gamma_\mu\gamma_5u\partial_\mu\pi$. In
this approach we find that one-$\pi$ exchange has no contribution to
the potential between two heavy vector mesons. Besides, it should be
noted that the flavor-SU(3) singlet and octet states of vector
mesons mix to form the physical $\omega$ and $\phi$ mesons, so the
exchange-mesons between two heavy mesons should not be the physical
mesons but rather the singlet and octet states. Then in the
interaction kernel between two heavy mesons SU(3) symmetry breaking
should be considered.

This paper has the structure as follows. In Sec. II the revised
general form of BS wave functions for the bound states composed of
two vector fields with arbitrary spin and definite parity is given.
In Sec. III we show the BS wave functions for the molecule states of
$D^{*0}\bar{D}^{*0}$ with $J^P=0^+$ and $2^+$. After constructing
the interaction kernel between two heavy vector mesons, we obtain
the Schr$\ddot{o}$dinger type equations in instantaneous
approximation. In Sec. IV we show how to calculate the form factors
of the heavy meson $D^*$. Then the interaction potential and the
mass of \emph{Y} are calculated. Our conclusion is presented in
Section V.

\section{REVISED GENERAL FORM OF THE BS WAVE FUNCTIONS}
\begin{figure}[!htb] \centering
\includegraphics[scale=1,width=9cm]{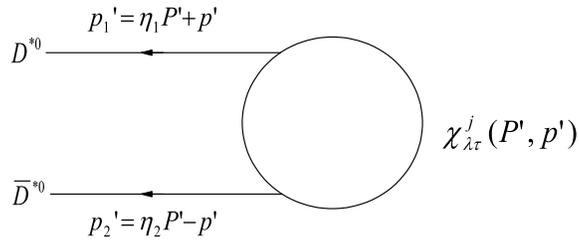}
\caption{\label{Fig0} Bethe-Salpeter wave function for the bound
state composed of two vector fields.}
\end{figure}

If a bound state of spin $j$ and parity $\eta_{P'}$ is composed of
two vector fields with masses $M_1$ and $M_2$, respectively, its BS
wave function is a $4\times4$ matrix
\begin{eqnarray}\label{BSwfdx}
\chi_{P'(\lambda\tau)}^j(x_1',x_2')=\langle0|TA_\lambda(x_1')A_\tau(x_2')|P',j\rangle,
\end{eqnarray}
which can be written as
\begin{eqnarray}
\chi_{P'(\lambda\tau)}^j(x_1',x_2')=e^{iP'\cdot
X'}\chi_{P'(\lambda\tau)}^j(x'),
\end{eqnarray}
where $X'=\eta_1x_1'+\eta_2x_2'$, $x'=x_1'-x_2'$ and $\eta_{1,2}$
are two positive quantities such that
$\eta_{1,2}=M_{1,2}/(M_1+M_2)$. Then one has the BS wave function in
the momentum representation
\begin{eqnarray}\label{BSwfdp}
\chi_{\lambda\tau}^j(P',p')=\int d^4x'e^{-ip'\cdot
x'}\chi_{P'(\lambda\tau)}^j(x'),
\end{eqnarray}
where $P'$ is the momentum of the bound state, $p'$ is the relative
momentum of two vector fields and we have $P'=p_1'+p_2',
p'=\eta_2p_1'-\eta_1p_2'$, $p_1'$ and $p_2'$ are the momenta of two
vector fields, respectively. This can be shown as Fig. \ref{Fig0}.

The polarization tensor $\eta_{\mu_1\mu_2\cdots\mu_j}$ describing
the spin of the bound state can be separated
\begin{eqnarray}\label{arbspin}
\chi_{\lambda\tau}^j(P',p')&=&\eta_{\mu_1\mu_2\cdots\mu_j}\chi_{\mu_1\mu_2\cdots\mu_j\lambda\tau}(P',p')
\end{eqnarray}
and the polarization tensor is totally symmetric, transverse and
traceless:
\begin{eqnarray}\label{polar}
\eta_{\mu_1\mu_2\cdots}=\eta_{\mu_2\mu_1\cdots},~~~~~~~P'_{\mu_1}\eta_{\mu_1\mu_2\cdots}=0,~~~~~~~\eta_{\mu_1\mu_2\cdots}^{\mu_1}=0.
\end{eqnarray}
Because of Eq. (\ref{polar}), $\chi_{\mu_1\cdots\mu_j\lambda\tau}$
is totally symmetric with respect to indices $\mu_1,\cdots,\mu_j$.
It is necessary to note that the polarization tensor
$\eta_{\mu_1\mu_2\cdots\mu_j}$ of the vector-vector bound state
should contain all contributions from the spins of two vector fields
and then $\chi_{\mu_1\cdots\mu_j\lambda\tau}$ should be independent
of the polarization vectors of two vector fields. Therefore, from
the BS wave function (\ref{BSwfdx}) and Lorentz covariance, we have
\begin{eqnarray}\label{scalarfun}
\chi_{\mu_1\cdots\mu_j\lambda\tau}&=&p'_{\mu_1}\cdots
p'_{\mu_j}[g_{\lambda\tau}f_1+(P'_\lambda p'_\tau+P'_\tau
p'_\lambda)f_2 +(P'_\lambda p'_\tau-P'_\tau
p'_\lambda)f_3+P'_\lambda P'_\tau f_4+p'_\lambda p'_\tau
f_5]\nonumber\\
&&+(p'_{\{\mu_2}\cdots p'_{\mu_j}g_{\mu_1\}\lambda}
p'_\tau+p'_{\{\mu_2}\cdots p'_{\mu_j}g_{\mu_1\}\tau} p'_\lambda)f_6\nonumber\\
&&+(p'_{\{\mu_2}\cdots p'_{\mu_j}g_{\mu_1\}\lambda}
p'_\tau-p'_{\{\mu_2}\cdots p'_{\mu_j}g_{\mu_1\}\tau} p'_\lambda)f_7\nonumber\\
&&+(p'_{\{\mu_2}\cdots p'_{\mu_j}g_{\mu_1\}\lambda}
P'_\tau+p'_{\{\mu_2}\cdots p'_{\mu_j}g_{\mu_1\}\tau} P'_\lambda)f_8\nonumber\\
&&+(p'_{\{\mu_2}\cdots p'_{\mu_j}g_{\mu_1\}\lambda}
P'_\tau-p'_{\{\mu_2}\cdots p'_{\mu_j}g_{\mu_1\}\tau} P'_\lambda)f_9\\
&&+p'_{\mu_1}\cdots p'_{\mu_j}\epsilon_{\lambda\tau\xi\zeta}p'_\xi
P'_\zeta f_{10}+p'_{\{\mu_2}\cdots
p'_{\mu_j}\epsilon_{\mu_1\}\lambda\tau\xi}p'_\xi
f_{11}+p'_{\{\mu_2}\cdots
p'_{\mu_j}\epsilon_{\mu_1\}\lambda\tau\xi}P'_\xi f_{12}\nonumber\\
&&+(p'_{\{\mu_2}\cdots
p'_{\mu_j}\epsilon_{\mu_1\}\lambda\xi\zeta}p'_\xi P'_\zeta
p'_\tau+p'_{\{\mu_2}\cdots
p'_{\mu_j}\epsilon_{\mu_1\}\tau\xi\zeta}p'_\xi P'_\zeta
p'_\lambda)f_{13}\nonumber\\
&&+(p'_{\{\mu_2}\cdots
p'_{\mu_j}\epsilon_{\mu_1\}\lambda\xi\zeta}p'_\xi P'_\zeta
p'_\tau-p'_{\{\mu_2}\cdots
p'_{\mu_j}\epsilon_{\mu_1\}\tau\xi\zeta}p'_\xi P'_\zeta
p'_\lambda)f_{14}\nonumber\\
&&+(p'_{\{\mu_2}\cdots
p'_{\mu_j}\epsilon_{\mu_1\}\lambda\xi\zeta}p'_\xi P'_\zeta
P'_\tau+p'_{\{\mu_2}\cdots
p'_{\mu_j}\epsilon_{\mu_1\}\tau\xi\zeta}p'_\xi P'_\zeta
P'_\lambda)f_{15}\nonumber\\
&&+(p'_{\{\mu_2}\cdots
p'_{\mu_j}\epsilon_{\mu_1\}\lambda\xi\zeta}p'_\xi P'_\zeta
P'_\tau-p'_{\{\mu_2}\cdots
p'_{\mu_j}\epsilon_{\mu_1\}\tau\xi\zeta}p'_\xi P'_\zeta
P'_\lambda)f_{16}\nonumber,
\end{eqnarray}
where $\{\mu_1,\cdots,\mu_j\}$ represents symmetrization of the
indices $\mu_1,\cdots,\mu_j$. There are only 16 scalar functions
$f_i(P'\cdot p',p'^2)(i=1,\cdots,16)$ in Eq. (\ref{scalarfun}). From
the massive vector field commutators for arbitrary times $x'_{10}$
and $x'_{20}$
\begin{eqnarray}
&&[A_\lambda(x'_1),A_\tau(x'_2)]=0~~~~\text{for}~~~M_1\neq
M_2,\nonumber\\
&&[A_\lambda(x'_1),A_\tau(x'_2)]=i(\delta_{\lambda\tau}-\frac{\partial_\lambda\partial_\tau}{M_1^2})\Delta'(x'_1-x'_2)~~~~\text{for}~~~M_1=M_2\nonumber,
\end{eqnarray}
where the right side of second equation is a c-number function, one
can have
\begin{eqnarray}
\langle0|TA_\lambda(x_1')A_\tau(x_2')|P',j\rangle=\langle0|TA_\tau(x_2')A_\lambda(x_1')|P',j\rangle.
\end{eqnarray}
So we obtain that in momentum representation the BS wave function of
two same or different vector fields is invariant under the
substitutions $p'_1\rightarrow p'_2$ and $p'_2\rightarrow p'_1$,
i.e.,
\begin{eqnarray}\label{BSwfinv}
\chi_{\lambda\tau}^j(P',p')=\chi_{\tau\lambda}^j(P',-p').
\end{eqnarray}
This invariance is similar to crossing symmetry which implies that
the scalar functions in Eq. (\ref{scalarfun}) have the following
properties: for $j=2n,~n=0,1,2,3\cdots$,
\begin{eqnarray}
f_i(P'\cdot p',p'^2)&=&+f_i(-P'\cdot
p',p'^2)~~~i=1,3,4,5,6,9,10,12,14,15,\nonumber\\
f_i(P'\cdot p',p'^2)&=&-f_i(-P'\cdot
p',p'^2)~~~i=2,7,8,11,13,16,\nonumber
\end{eqnarray}
and, for $j=2n+1,~n=0,1,2,3\cdots$,
\begin{eqnarray}
f_i(P'\cdot p',p'^2)&=&+f_i(-P'\cdot
p',p'^2)~~~i=2,7,8,11,13,16,\nonumber\\
f_i(P'\cdot p',p'^2)&=&-f_i(-P'\cdot
p',p'^2)~~~i=1,3,4,5,6,9,10,12,14,15.
\end{eqnarray}
Then we will reduce the general form without any assumption and
approximation.

Using the subsidiary condition $\partial_\mu A_\mu(x)=0$ for the
massive vector field and the equal-time commutation relation, we get
\begin{eqnarray}
\partial_{1\lambda}TA_\lambda(x_1')A_\tau(x_2')=\partial_{2\tau}TA_\lambda(x_1')A_\tau(x_2')=0\nonumber.
\end{eqnarray}
The proof has been given by Ref. \cite{mypaper3}. The BS wave
function in Eq. (\ref{BSwfdx}) has this relation
\begin{eqnarray}
\partial_{1\lambda}\langle0|TA_\lambda(x_1')A_\tau(x_2')|P',j\rangle=\partial_{2\tau}\langle0|TA_\lambda(x_1')A_\tau(x_2')|P',j\rangle
=0,
\end{eqnarray}
and in momentum representation
\begin{eqnarray}\label{proofvec}
p'_{1\lambda}\chi^j_{\lambda\tau}(P',p')=p'_{2\tau}\chi^j_{\lambda\tau}(P',p')=0.
\end{eqnarray}

To satisfy Eqs. (\ref{BSwfinv}) and (\ref{proofvec}), the BS wave
function of the bound state becomes
\begin{eqnarray}\label{BSwfj0}
\chi_{\lambda\tau}^{j=0}(P',p')&=&[(p_1'\cdot
p_2')g_{\lambda\tau}-p'_{2\lambda}p'_{1\tau}]\phi_1+[p_1'^2p_2'^2g_{\lambda\tau}+(p_1'\cdot
p_2')p'_{1\lambda}p'_{2\tau}\nonumber\\
&&-p_2'^2p'_{1\lambda}p'_{1\tau}-p_1'^2p'_{2\lambda}p'_{2\tau}]\phi_2+\epsilon_{\lambda\tau\xi\zeta}p'_\xi
P'_\zeta\psi_1,
\end{eqnarray}
\begin{eqnarray}\label{BSwfj}
&&\chi_{\lambda\tau}^{j\neq0}(P',p')=\nonumber\\
&&\eta_{\mu_1\cdots\mu_j}\{p'_{\mu_1}\cdots p'_{\mu_j}\{[(p_1'\cdot
p_2')g_{\lambda\tau}-p'_{2\lambda}p'_{1\tau}]\phi_1\nonumber\\
&&+[p_1'^2p_2'^2g_{\lambda\tau}+(p_1'\cdot
p_2')p'_{1\lambda}p'_{2\tau}-p_2'^2p'_{1\lambda}p'_{1\tau}-p_1'^2p'_{2\lambda}p'_{2\tau}]\phi_2\}\nonumber\\
&&+\{\frac{1}{j!}p'_{\{\mu_2}\cdots
p'_{\mu_j}g_{\mu_1\}\lambda}[p'^2_1p'^2_2p'_{1\tau}-(p'_1\cdot
p'_2)p'^2_1p'_{2\tau}]-p'_{\mu_1}\cdots
p'_{\mu_j}[p'^2_2p'_{1\lambda} p'_{1\tau}-(p'_1\cdot
p'_2)p'_{1\lambda}
p'_{2\tau}]\}\phi_3\nonumber\\
&&+\{\frac{1}{j!}p'_{\{\mu_2}\cdots
p'_{\mu_j}g_{\mu_1\}\tau}[(p'_1\cdot
p'_2)p'^2_2p'_{1\lambda}-p'^2_1p'^2_2p'_{2\lambda}]+p'_{\mu_1}\cdots
p'_{\mu_j}[(p'_1\cdot p'_2)p'_{1\lambda}
p'_{2\tau}-p'^2_1p'_{2\lambda} p'_{2\tau}]\}\phi_4\nonumber\\
&&+p'_{\mu_1}\cdots
p'_{\mu_j}\epsilon_{\lambda\tau\xi\zeta}p'_\xi P'_\zeta\psi_1\\
&&+\{[(p'_2\cdot p')-(p'_1\cdot p')]p'_{\{\mu_2}\cdots
p'_{\mu_j}\epsilon_{\mu_1\}\lambda\tau\xi}p'_\xi+[\eta_2(p'_1\cdot
p')+\eta_1(p'_2\cdot p')]p'_{\{\mu_2}\cdots
p'_{\mu_j}\epsilon_{\mu_1\}\lambda\tau\xi}P'_\xi\nonumber\\
&&+p'_{\{\mu_2}\cdots
p'_{\mu_j}\epsilon_{\mu_1\}\lambda\xi\zeta}p'_\xi P'_\zeta
p'_\tau+p'_{\{\mu_2}\cdots
p'_{\mu_j}\epsilon_{\mu_1\}\tau\xi\zeta}p'_\xi P'_\zeta
p'_\lambda\}\psi_2\nonumber\\
&&+\{[(p'_1\cdot p')+(p'_2\cdot p')]p'_{\{\mu_2}\cdots
p'_{\mu_j}\epsilon_{\mu_1\}\lambda\tau\xi}p'_\xi+[\eta_1(p'_2\cdot
p')-\eta_2(p'_1\cdot p')]p'_{\{\mu_2}\cdots
p'_{\mu_j}\epsilon_{\mu_1\}\lambda\tau\xi}P'_\xi\nonumber\\
&&+p'_{\{\mu_2}\cdots
p'_{\mu_j}\epsilon_{\mu_1\}\lambda\xi\zeta}p'_\xi P'_\zeta
p'_\tau-p'_{\{\mu_2}\cdots
p'_{\mu_j}\epsilon_{\mu_1\}\tau\xi\zeta}p'_\xi P'_\zeta
p'_\lambda\}\psi_3\nonumber\\
&&+\{[(p'_2\cdot P')-(p'_1\cdot P')]p'_{\{\mu_2}\cdots
p'_{\mu_j}\epsilon_{\mu_1\}\lambda\tau\xi}p'_\xi+[\eta_2(p'_1\cdot
P')+\eta_1(p'_2\cdot P')]p'_{\{\mu_2}\cdots
p'_{\mu_j}\epsilon_{\mu_1\}\lambda\tau\xi}P'_\xi\nonumber\\
&&+p'_{\{\mu_2}\cdots
p'_{\mu_j}\epsilon_{\mu_1\}\lambda\xi\zeta}p'_\xi P'_\zeta
P'_\tau+p'_{\{\mu_2}\cdots
p'_{\mu_j}\epsilon_{\mu_1\}\tau\xi\zeta}p'_\xi P'_\zeta
P'_\lambda\}\psi_4\nonumber\\
&&+\{[(p'_1\cdot P')+(p'_2\cdot P')]p'_{\{\mu_2}\cdots
p'_{\mu_j}\epsilon_{\mu_1\}\lambda\tau\xi}p'_\xi+[\eta_1(p'_2\cdot
P')-\eta_2(p'_1\cdot P')]p'_{\{\mu_2}\cdots
p'_{\mu_j}\epsilon_{\mu_1\}\lambda\tau\xi}P'_\xi\nonumber\\
&&+p'_{\{\mu_2}\cdots
p'_{\mu_j}\epsilon_{\mu_1\}\lambda\xi\zeta}p'_\xi P'_\zeta
P'_\tau-p'_{\{\mu_2}\cdots
p'_{\mu_j}\epsilon_{\mu_1\}\tau\xi\zeta}p'_\xi P'_\zeta
P'_\lambda\}\psi_5\nonumber\},
\end{eqnarray}
where $\phi_i$ and $\psi_i$ are the linear combinations of
$f_i(i=1,\cdots, 16)$ in Eq. (\ref{scalarfun}), respectively. This
derivation makes use of the fact that $P'$, $p'$ and
$\eta_{\mu_1\cdots\mu_j}$ are linearly independent.

Now, under space reflection
\begin{eqnarray}
&&x_i'\rightarrow x_i''=-x_i',~~x_0''=x_0',\nonumber
\end{eqnarray}
one has
\begin{eqnarray}\label{spaceref}
&&\textsf{P}A_\lambda(x')\textsf{P}^{-1}=\mathscr{P}_{\lambda\xi}
A_\xi(x''),~~~~~~\mathscr{P}_{\lambda\xi}=\left(\begin{array}{cccc}
-1&0&0&0\\0&-1&0&0\\0&0&-1&0\\0&0&0&1
\end{array}\right),\nonumber\\
&&\textsf{P}|0\rangle=|0\rangle,~~~~~~~\textsf{P}|P',j\rangle=\eta_{P'}|P'',j\rangle,
\end{eqnarray}
with $x''=(-\textbf{x}',x_0')$ and $P''=(-\textbf{P}',P_0')$.

We obtain the properties of the BS wave function under space
reflection from Eqs. (\ref{BSwfdx}) and (\ref{spaceref}):
\begin{eqnarray}
\chi_{P'(\lambda\tau)}^{j}(x_1',x_2')=\eta_{P'}\mathscr{P}_{\lambda\xi}\mathscr{P}_{\tau\zeta}\chi_{P''(\xi\zeta)}^{j}(x_1'',x_2'')
\end{eqnarray}
and in the momentum representation
\begin{eqnarray}\label{momspaceref}
\chi_{\lambda\tau}^{j}(P',p')=\eta_{P'}\mathscr{P}_{\lambda\xi}\mathscr{P}_{\tau\zeta}\chi_{\xi\zeta}^{j}(P'',p'').
\end{eqnarray}

From Eqs. (\ref{arbspin}), (\ref{scalarfun}), (\ref{BSwfj0}),
(\ref{BSwfj}) and (\ref{momspaceref}), it is easy to derive that,
for $\eta_{P'}=(-1)^j$,
\begin{eqnarray}\label{jp0}
\chi_{\lambda\tau}^{j=0}(P',p')&=&[(p_1'\cdot
p_2')g_{\lambda\tau}-p'_{2\lambda}p'_{1\tau}]\phi_1\nonumber\\
&&+[p_1'^2p_2'^2g_{\lambda\tau}+(p_1'\cdot
p_2')p'_{1\lambda}p'_{2\tau}-p_2'^2p'_{1\lambda}p'_{1\tau}-p_1'^2p'_{2\lambda}p'_{2\tau}]\phi_2,
\end{eqnarray}
\begin{eqnarray}\label{jp}
\chi_{\lambda\tau}^{j\neq0}(P',p')&=&\eta_{\mu_1\cdots\mu_j}\{p'_{\mu_1}\cdots
p'_{\mu_j}\{[(p_1'\cdot
p_2')g_{\lambda\tau}-p'_{2\lambda}p'_{1\tau}]\phi_1\nonumber\\
&&+[p_1'^2p_2'^2g_{\lambda\tau}+(p_1'\cdot
p_2')p'_{1\lambda}p'_{2\tau}-p_2'^2p'_{1\lambda}p'_{1\tau}-p_1'^2p'_{2\lambda}p'_{2\tau}]\phi_2\}\nonumber\\
&&+\{\frac{1}{j!}p'_{\{\mu_2}\cdots
p'_{\mu_j}g_{\mu_1\}\lambda}[p'^2_1p'^2_2p'_{1\tau}-(p'_1\cdot
p'_2)p'^2_1p'_{2\tau}]\nonumber\\&&
-p'_{\mu_1}\cdots
p'_{\mu_j}[p'^2_2p'_{1\lambda} p'_{1\tau}-(p'_1\cdot
p'_2)p'_{1\lambda}
p'_{2\tau}]\}\phi_3\nonumber\\
&&+\{\frac{1}{j!}p'_{\{\mu_2}\cdots
p'_{\mu_j}g_{\mu_1\}\tau}[(p'_1\cdot
p'_2)p'^2_2p'_{1\lambda}-p'^2_1p'^2_2p'_{2\lambda}]\nonumber\\
&&+p'_{\mu_1}\cdots p'_{\mu_j}[(p'_1\cdot p'_2)p'_{1\lambda}
p'_{2\tau}-p'^2_1p'_{2\lambda} p'_{2\tau}]\}\phi_4\},
\end{eqnarray}
and, for $\eta_{P'}=(-1)^{j+1}$,
\begin{eqnarray}\label{jm0}
\chi_{\lambda\tau}^{j=0}(P',p')=\epsilon_{\lambda\tau\xi\zeta}p'_\xi
P'_\zeta\psi_1,
\end{eqnarray}
\begin{eqnarray}\label{jm}
&&\chi_{\lambda\tau}^{j\neq0}(P',p')=\eta_{\mu_1\cdots\mu_j}\{p'_{\mu_1}\cdots
p'_{\mu_j}\epsilon_{\lambda\tau\xi\zeta}p'_\xi P'_\zeta\psi_1\\
&&+\{[(p'_2\cdot p')-(p'_1\cdot p')]p'_{\{\mu_2}\cdots
p'_{\mu_j}\epsilon_{\mu_1\}\lambda\tau\xi}p'_\xi+[\eta_2(p'_1\cdot
p')+\eta_1(p'_2\cdot p')]p'_{\{\mu_2}\cdots
p'_{\mu_j}\epsilon_{\mu_1\}\lambda\tau\xi}P'_\xi\nonumber\\
&&+p'_{\{\mu_2}\cdots
p'_{\mu_j}\epsilon_{\mu_1\}\lambda\xi\zeta}p'_\xi P'_\zeta
p'_\tau+p'_{\{\mu_2}\cdots
p'_{\mu_j}\epsilon_{\mu_1\}\tau\xi\zeta}p'_\xi P'_\zeta
p'_\lambda\}\psi_2\nonumber\\
&&+\{[(p'_1\cdot p')+(p'_2\cdot p')]p'_{\{\mu_2}\cdots
p'_{\mu_j}\epsilon_{\mu_1\}\lambda\tau\xi}p'_\xi+[\eta_1(p'_2\cdot
p')-\eta_2(p'_1\cdot p')]p'_{\{\mu_2}\cdots
p'_{\mu_j}\epsilon_{\mu_1\}\lambda\tau\xi}P'_\xi\nonumber\\
&&+p'_{\{\mu_2}\cdots
p'_{\mu_j}\epsilon_{\mu_1\}\lambda\xi\zeta}p'_\xi P'_\zeta
p'_\tau-p'_{\{\mu_2}\cdots
p'_{\mu_j}\epsilon_{\mu_1\}\tau\xi\zeta}p'_\xi P'_\zeta
p'_\lambda\}\psi_3\nonumber\\
&&+\{[(p'_2\cdot P')-(p'_1\cdot P')]p'_{\{\mu_2}\cdots
p'_{\mu_j}\epsilon_{\mu_1\}\lambda\tau\xi}p'_\xi+[\eta_2(p'_1\cdot
P')+\eta_1(p'_2\cdot P')]p'_{\{\mu_2}\cdots
p'_{\mu_j}\epsilon_{\mu_1\}\lambda\tau\xi}P'_\xi\nonumber\\
&&+p'_{\{\mu_2}\cdots
p'_{\mu_j}\epsilon_{\mu_1\}\lambda\xi\zeta}p'_\xi P'_\zeta
P'_\tau+p'_{\{\mu_2}\cdots
p'_{\mu_j}\epsilon_{\mu_1\}\tau\xi\zeta}p'_\xi P'_\zeta
P'_\lambda\}\psi_4\nonumber\\
&&+\{[(p'_1\cdot P')+(p'_2\cdot P')]p'_{\{\mu_2}\cdots
p'_{\mu_j}\epsilon_{\mu_1\}\lambda\tau\xi}p'_\xi+[\eta_1(p'_2\cdot
P')-\eta_2(p'_1\cdot P')]p'_{\{\mu_2}\cdots
p'_{\mu_j}\epsilon_{\mu_1\}\lambda\tau\xi}P'_\xi\nonumber\\
&&+p'_{\{\mu_2}\cdots
p'_{\mu_j}\epsilon_{\mu_1\}\lambda\xi\zeta}p'_\xi P'_\zeta
P'_\tau-p'_{\{\mu_2}\cdots
p'_{\mu_j}\epsilon_{\mu_1\}\tau\xi\zeta}p'_\xi P'_\zeta
P'_\lambda\}\psi_5\nonumber\}.
\end{eqnarray}

The general form of the BS wave functions for the bound states
composed of two massive vector fields of arbitrary spin and definite
parity is obtained. No matter how high the spin of the bound state
is, its BS wave function should satisfy Eq. (\ref{jp0}), (\ref{jp}),
(\ref{jm0}) or (\ref{jm}). From Eqs. (\ref{jp0}) and (\ref{jp}), we
conclude that the BS wave function of a bound state composed of two
massive vector fields with spin $j\neq0$ and parity $(-1)^j$ has
only four independent components and that of a bound state with
$J^P=0^+$ has only two independent components. From Eqs. (\ref{jm0})
and (\ref{jm}), we conclude that the BS wave function of a bound
state with spin $j$ and parity $(-1)^{j+1}$ has only five
independent components except for $j=0$ and one for $j=0$. Up to
now, all the above analyses are model independent. In the next
section we will apply the general formalism to investigate molecule
states composed of two vector mesons.

\section{THE EXTENDED BETHE-SALPETER EQUATION}
Assuming that the \emph{Y}(3940) is a S-wave molecule state
consisting of two heavy vector mesons $D^{*0}$ and $\bar{D}^{*0}$,
one can have $J^P=0^+$ or $2^+$ for this system. \cite{liu} From
Eqs. (\ref{jp0}) and (\ref{jp}), we can obtain the BS wave function
describing this bound state, for $J^P=0^+$,
\begin{eqnarray}\label{SBSWF}
\chi_{\lambda\tau}^{0^+}(P',p')&=&[(p_1'\cdot p_2')g_{\lambda\tau}
-p'_{2\lambda}p'_{1\tau}]\mathcal
{F}_1\nonumber\\
&&+[p_1'^2p_2'^2g_{\lambda\tau}+(p_1'\cdot
p_2')p'_{1\lambda}p'_{2\tau}-p_2'^2p'_{1\lambda}p'_{1\tau}-p_1'^2p'_{2\lambda}p'_{2\tau}]\mathcal
{F}_2,
\end{eqnarray}
or, for $J^P=2^+$,
\begin{eqnarray}\label{TBSWF}
\chi_{\lambda\tau}^{2^+}(P',p')&=&\eta_{\mu_1\mu_2}\{p'_{\mu_1}
p'_{\mu_2}\{[(p_1'\cdot p_2')g_{\lambda\tau}-p'_{2\lambda}p'_{1\tau}]\mathcal {G}_1\nonumber\\
&&+[p_1'^2p_2'^2g_{\lambda\tau}+(p_1'\cdot
p_2')p'_{1\lambda}p'_{2\tau}-p_2'^2p'_{1\lambda}p'_{1\tau}-p_1'^2p'_{2\lambda}p'_{2\tau}]\mathcal{G}_2\}\nonumber\\
&&+\{\frac{1}{2!}p'_{\{\mu_2}g_{\mu_1\}\lambda}[p'^2_1p'^2_2p'_{1\tau}-(p'_1\cdot
p'_2)p'^2_1p'_{2\tau}]\nonumber\\
&&-p'_{\mu_1}p'_{\mu_2}[p'^2_2p'_{1\lambda} p'_{1\tau}-(p'_1\cdot
p'_2)p'_{1\lambda}
p'_{2\tau}]\}\mathcal{G}_3\nonumber\\
&&+\{\frac{1}{2!}p'_{\{\mu_2}g_{\mu_1\}\tau}[(p'_1\cdot
p'_2)p'^2_2p'_{1\lambda}-p'^2_1p'^2_2p'_{2\lambda}]\nonumber\\
&&+p'_{\mu_1}p'_{\mu_2}[(p'_1\cdot p'_2)p'_{1\lambda}
p'_{2\tau}-p'^2_1p'_{2\lambda} p'_{2\tau}]\}\mathcal{G}_4\}.
\end{eqnarray}

\begin{figure}[!htb]
\centering
\includegraphics[scale=1,width=8cm]{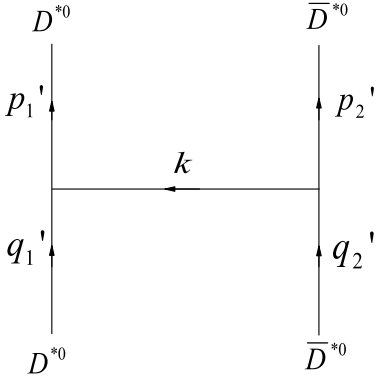}
\caption{\label{Fig1} The light meson exchange between two heavy
mesons.}
\end{figure}

The BS wave function of this bound state satisfies the
Bethe-Salpeter equation
\begin{eqnarray}\label{BSE1}
\chi_{\lambda\tau}(P',p')=\int
\frac{id^4q'}{(2\pi)^4}\Delta_{F\lambda\alpha}(p_1')\mathcal
{V}_{\alpha\theta,\beta\kappa}(p',q';P')\chi_{\theta\kappa}(P',q')\Delta_{F\beta\tau}(p_2'),
\end{eqnarray}
where $\mathcal{V}_{\alpha\theta,\beta\kappa}$ is the interaction
kernel, the propagators for the spin 1 fields
$\Delta_{F\lambda\alpha}(p_1')=(\delta_{\lambda\alpha}+\frac{p'_{1\lambda}
p'_{1\alpha}}{M_1^2})\frac{1}{p_1'^2+M_1^2-i\epsilon}$,
$\Delta_{F\beta\tau}(p_2')=(\delta_{\beta\tau}+\frac{p'_{2\beta}
p'_{2\tau}}{M_2^2})\frac{1}{p_2'^2+M_2^2-i\epsilon}$ and the bound
state momentum is set as $P'=(0,0,0,iM)$ in the rest frame. In Ref.
\cite{mypaper3}, we have considered that the effective interaction
between two heavy mesons is derived from one light meson ($\sigma$,
$\omega$, $\rho$) exchange and obtained that the molecule state
$D^{*0}\bar D^{*0}$ lies above the threshold. In this work,
one-$\pi$ exchange is also considered and one-$\omega$ exchange is
reconsidered, shown as in Fig. \ref{Fig1}.

Now, we construct the kernel between two heavy vector mesons from
one-$\pi$ exchange. The charmed meson $D^{*}$ is composed of a heavy
quark $c$ and a light antiquark $\bar u$. Owing to the large mass of
$c$-quark, the Lagrangian representing the interaction of
$\pi$-meson triplet with quarks should be
\begin{eqnarray}
\mathscr{L}_I=ig_\pi\left(\begin{array}{ccc} \bar{u}&\bar{d}&\bar{c}
\end{array}\right)\gamma_5\left(\begin{array}{ccc}
\pi^0&\sqrt{2}\pi^+&0\\\sqrt{2}\pi^-&-\pi^0&0\\0&0&0
\end{array}\right)\left(\begin{array}{c} u\\d\\c
\end{array}\right),
\end{eqnarray}
or
\begin{eqnarray}
\mathscr{L}_I=if_\pi\left(\begin{array}{ccc} \bar{u}&\bar{d}&\bar{c}
\end{array}\right)\gamma_\mu\gamma_5\left(\begin{array}{ccc}
\partial_\mu\pi^0&\sqrt{2}\partial_\mu\pi^+&0\\\sqrt{2}\partial_\mu\pi^-&-\partial_\mu\pi^0&0\\0&0&0
\end{array}\right)\left(\begin{array}{c} u\\d\\c
\end{array}\right),
\end{eqnarray}
where $g_\pi$ and $f_\pi$ are the $\pi$-meson-quark coupling
constants, $g_\pi=\frac{340}{97}$ \cite{cc1a,cc1b}. Because the
contribution of Fig. \ref{Fig1} is from the term
$i\bar{u}\gamma_5\pi^0u$ or
$i\bar{u}\gamma_\mu\gamma_5\partial_\mu\pi^0u$, we set
$f_\pi=\frac{g_\pi}{2m_u}$ and $m_u$ is u-quark mass. Then the
effective quark current is $J^-=i\bar{u}\gamma_5 u$ or
$J^+_\mu=i\bar{u}\gamma_\mu\gamma_5 u$ and the S-matrix element
between two heavy mesons is
\begin{eqnarray}\label{Vpse1}
V_\pi=g_\pi^2\langle
V(p_1')|J^-|V(q_1')\rangle\frac{1}{k^2+m_\pi^2}\langle
V(p_2')|J^-|V(q_2')\rangle,
\end{eqnarray}
or
\begin{eqnarray}\label{Vpse}
V_\pi=f_\pi^2\langle V(p_1')|J^+_\mu|V(q_1')\rangle
k_\mu\frac{1}{k^2+m_\pi^2}k_\nu\langle
V(p_2')|J^+_\nu|V(q_2')\rangle,
\end{eqnarray}
where $\langle V|J^-|V\rangle$ and $\langle V|J^+_\mu|V\rangle$
represent the vertices of the pseudoscalar meson interaction with
the heavy vector meson, respectively. The matrix elements of these
quark currents can be expressed as
\begin{eqnarray}\label{meofqc1}
\langle V(p_1')|J^-|V(q_1')
\rangle&=&\frac{1}{\sqrt{2E_1(p_1')}\sqrt{2E_1(q_1')}}h^{-}(k^2)\epsilon_{abcd}p_{1a}'q_{1b}'\varepsilon^*_c(p_1')
\varepsilon_d(q_1'),
\end{eqnarray}
\begin{eqnarray}\label{meofqc2}
\langle V(p_2')|J^-|V(q_2')
\rangle&=&\frac{1}{2\sqrt{E_2(p_2')E_2(q_2')}}\bar
h^{-}(k^2)\epsilon_{a'b'c'd'}p_{2a'}'q_{2b'}'\varepsilon^*_{c'}(p_2')
\varepsilon_{d'}(q_2'),
\end{eqnarray}
\begin{eqnarray}\label{meofqc3}
\langle V(p_1')|J^+_{\mu}|V(q_1')
\rangle&=&\frac{1}{2\sqrt{E_1(p_1')E_1(q_1')}}h^{(p)}(k^2)\epsilon_{\mu
abc}\varepsilon^*_a(p_1') \varepsilon_b(q_1')(p_{1}'+q_{1}')_c,
\end{eqnarray}
\begin{eqnarray}\label{meofqc4}
\langle V(p_2')|J^+_{\nu}|V(q_2')
\rangle&=&\frac{1}{2\sqrt{E_2(p_2')E_2(q_2')}}\bar
h^{(p)}(k^2)\epsilon_{\nu a'b'c'}\varepsilon^*_{a'}(p_2')
\varepsilon_{b'}(q_2')(p_{2}'+q_{2}')_{c'},
\end{eqnarray}
where $p'_1=(\textbf{p}',ip_{10}')$, $p'_2=(-\textbf{p}',ip_{20}')$,
$q'_1=(\textbf{q}',iq_{10}')$, $q'_2=(-\textbf{q}',iq_{20}')$,
$k=p_1'-q_1'=q_2'-p_2'$ is the momentum of the light meson and
$\textbf{k}=\textbf{p}'-\textbf{q}'$; $h(k^2)$ and $\bar h(k^2)$ are
scalar functions, the four-vector
$\varepsilon(p)=(\bm{\varepsilon}+\frac{(\bm
{\varepsilon}\cdot\textbf{p})\textbf{p}}{M_H(E_H(p)+M_H)},i\frac{\bm{\varepsilon}\cdot\textbf{p}}{M_H})$
is the polarization vector of heavy vector meson with momentum
$\text{p}$, $E_H(p)=\sqrt{\textbf{p}^2+M_H^2}$,
$(\bm{\varepsilon},0)$ is the polarization vector in the heavy meson
rest frame. In our approach, considering that the exchange-meson is
off the mass shell, we calculate the meson-meson interaction when
$k^2\neq -m^2$ and the heavy meson form factors $h(k^2)$ and $\bar
h(k^2)$ are necessarily required. In Sec. IV, we will show that the
form factors $h^{-}(k^2)=\bar h^{-}(k^2)=0$. Then the effective
interaction from one-$\pi$ exchange should have the form given by
Eq. (\ref{Vpse}). Cutting the external lines containing the
normalizations and polarization vectors $\varepsilon^*_\alpha(p_1'),
\varepsilon_\theta(q_1'), \varepsilon^*_\beta(p_2'),
\varepsilon_{\kappa}(q_2')$, we obtain the interaction kernel from
one-$\pi$ exchange
\begin{eqnarray}
\mathcal
{V}^\pi_{\alpha\theta,\beta\kappa}(p',q';P')&=&f_\pi^2h_1^{(p)}(k^2)\epsilon_{\mu\alpha\theta
c}(p_1'+q_1')_ck_\mu\frac{1}{k^2+m_\pi^2}\bar{h}_1^{(p)}(k^2)\epsilon_{\nu\beta\kappa
c'}(p_2'+q_2')_{c'}k_\nu.
\end{eqnarray}

Then we reconsider one-$\omega$ exchange between two heavy vector
mesons. In hadronic physics, the physical $\omega$ and $\phi$ mesons
are linear combinations of the SU(3) octet $V_8=(u\bar u+d\bar
d-2s\bar s)/\sqrt6$ and singlet $V_1=(u\bar u+d\bar d+s\bar
s)/\sqrt3$ as
\begin{eqnarray}\label{mix}
\phi=-V_8cos\theta+V_1sin\theta,~~~~~
\omega=V_8sin\theta+V_1cos\theta,
\end{eqnarray}
where the mixing angle $\theta=38.58^\circ$ obtained by
KLOE\cite{mix}. Because of the SU(3) symmetry, we consider that the
exchange-mesons are not the physical mesons and the exchanged mesons
should be the octet $V_8$ and singlet $V_1$ states. From Eq.
(\ref{mix}), we can obtain the relation of the octet-quark coupling
constant $g_8$ and the singlet-quark coupling constant $g_1$
\begin{eqnarray}
g_\phi=-g_8cos\theta+g_1sin\theta,~~~~~
g_\omega=g_8sin\theta+g_1cos\theta,
\end{eqnarray}
where $g_\omega$ and $g_\phi$ are the corresponding meson-quark
coupling constants, $g_\omega^2=2.42$ and $g_\phi^2=13.0$
\cite{cc2}. Since the SU(3) is broken, the masses of the singlet
$V_1$ and octet $V_8$ states are approximatively identified with the
two physical masses of $\omega$ and $\phi$ mesons, respectively. The
interaction kernel derived from one scalar meson exchange and one
vector meson exchange has been given in Ref. \cite{mypaper3}, and
the kernel from one light meson ($\pi$, $\sigma$, $\rho$, $V_1$ and
$V_8$) exchange becomes
\begin{eqnarray}\label{kernel}
&&\mathcal {V}_{\alpha\theta,\beta\kappa}(p',q';P')=
h_1^{(p)}(k^2)\epsilon_{\mu\alpha\theta
c}(p_1'+q_1')_ck_\mu\frac{f_\pi^2}{k^2+m_\pi^2}\bar{h}_1^{(p)}(k^2)\epsilon_{\nu\beta\kappa
c'}(p_2'+q_2')_{c'}k_\nu\nonumber\\
&&+h_1^{(s)}(k^2)\frac{g_\sigma^2}{k^2+m_\sigma^2}\bar{h}_1^{(s)}(k^2)\delta_{\alpha\theta}\delta_{\beta\kappa}+(\frac{g_\rho^2}{k^2+m_\rho^2}+\frac{g_1^2}{k^2+m_\omega^2}+\frac{g_8^2}{k^2+m_\phi^2})\{h_1^{(\text{v})}(k^2)\bar{h}_1^{(\text{v})}(k^2)\nonumber\\
&&\times(p_1'+q_1')\cdot(p_2'+q_2')\delta_{\alpha\theta}\delta_{\beta\kappa}-h_1^{(\text{v})}(k^2)\bar{h}_2^{(\text{v})}(k^2)\delta_{\alpha\theta}[q_{2\beta}'(p_1'+q_1')_\kappa+(p_1'+q_1')_\beta p_{2\kappa}']\nonumber\\
&&-h_2^{(\text{v})}(k^2)\bar{h}_1^{(\text{v})}(k^2)[q_{1\alpha}'(p_2'+q_2')_\theta+(p_2'+q_2')_\alpha
p_{1\theta}']\delta_{\beta\kappa}+h_2^{(\text{v})}(k^2)\bar{h}_2^{(\text{v})}(k^2)[q_{1\alpha}'q_{2\beta}'\delta_{\theta\kappa}\nonumber\\
&&+q_{1\alpha}'\delta_{\theta\beta}p_{2\kappa}'+\delta_{\alpha\kappa}p_{1\theta}'q_{2\beta}'+\delta_{\alpha\beta}p_{1\theta}'p_{2\kappa}']\},
\end{eqnarray}
where $k=(\textbf{k},0)$.

Firstly, we assume that the \emph{Y}(3940) is a molecule state with
$J^P=0^+$. Substituting its BS wave function given by Eq.
(\ref{SBSWF}) and the kernel (\ref{kernel}) into the BS equation
(\ref{BSE1}), we find that the integral of one term on the
right-hand side of Eq. (\ref{SBSWF}) has contribution to the one of
itself and the other term. Moreover, the cross terms contain the
factors of $\frac{1}{M_1^2}$ and $\frac{1}{M_2^2}$, which are small
for the masses of the heavy mesons are large. It is difficult to
strictly solve the BS equation, in this paper we use a simple
approach to solve it as follow. Ignoring the cross terms, one can
obtain two individual equations:
\begin{eqnarray}\label{BSE2}
&&\mathcal{F}^1_{\lambda\tau}(P'\cdot p',p'^2) =\int
\frac{id^4q'}{(2\pi)^4}\Delta_{F\lambda\alpha}(p_1')\mathcal
{V}_{\alpha\theta,\beta\kappa}(p',q';P')\mathcal{F}^1_{\theta\kappa}(P'\cdot
q',q'^2)\Delta_{F\beta\tau}(p_2'),
\end{eqnarray}
\begin{eqnarray}\label{BSE3}
&&\mathcal {F}^2_{\lambda\tau}(P'\cdot p',p'^2) =\int
\frac{id^4q'}{(2\pi)^4}\Delta_{F\lambda\alpha}(p_1')\mathcal
{V}_{\alpha\theta,\beta\kappa}(p',q';P')\mathcal{F}^2_{\theta\kappa}(P'\cdot
q',q'^2)\Delta_{F\beta\tau}(p_2'),
\end{eqnarray}
where $\mathcal{F}^1_{\lambda\tau}(P'\cdot p',p'^2)=[(p_1'\cdot
p_2')g_{\lambda\tau}-p'_{2\lambda}p'_{1\tau}]\mathcal{F}_1(P'\cdot
p',p'^2)$ and $\mathcal{F}^2_{\lambda\tau}(P'\cdot
p',p'^2)=[p_1'^2p_2'^2g_{\lambda\tau}+(p_1'\cdot
p_2')p'_{1\lambda}p'_{2\tau}-p_2'^2p'_{1\lambda}p'_{1\tau}-p_1'^2p'_{2\lambda}p'_{2\tau}]\mathcal
{F}_2(P'\cdot p',p'^2)$. And solving this two equations
respectively, one can obtain two series of eigenvalues and
eigenfunctions. Because the cross terms are small, we can take the
ground state BS wave function to be a linear combination of two
eigenstates $\mathcal{F}^{10}_{\lambda\tau}$ and
$\mathcal{F}^{20}_{\lambda\tau}$ corresponding to lowest energy in
Eqs. (\ref{BSE2}) and (\ref{BSE3}). Then in the basis provided by
$\mathcal{F}^{10}_{\lambda\tau}=[(p_1'\cdot
p_2')g_{\lambda\tau}-p'_{2\lambda}p'_{1\tau}]\mathcal{F}_{10}(P'\cdot
p',p'^2)$ and
$\mathcal{F}^{20}_{\lambda\tau}=[p_1'^2p_2'^2g_{\lambda\tau}+(p_1'\cdot
p_2')p'_{1\lambda}p'_{2\tau}-p_2'^2p'_{1\lambda}p'_{1\tau}-p_1'^2p'_{2\lambda}p'_{2\tau}]\mathcal
{F}_{20}(P'\cdot p',p'^2)$, the BS wave function
$\chi^{0^+}_{\lambda\tau}$ is considered as
\begin{eqnarray}\label{BSwfapprox}
\chi^{0^+}_{\lambda\tau}(P',p')=c_1\mathcal{F}^{10}_{\lambda\tau}(P'\cdot
p',p'^2)+c_2\mathcal{F}^{20}_{\lambda\tau}(P'\cdot p',p'^2).
\end{eqnarray}
Substituting Eq. (\ref{BSwfapprox}) into the BS equation
(\ref{BSE1}) and comparing the tensor structures in both sides, we
obtain a eigenvalue equation
\begin{eqnarray}\label{eigeneq}
&&c_1\mathcal{F}_{10}(P'\cdot
p',p'^2)=\nonumber\\
&&\frac{1}{p_1'^2+M_1^2-i\epsilon}\frac{1}{p_2'^2+M_2^2-i\epsilon}\{\int\frac{id^4q'}{(2\pi)^4}\{h_1^{(s)}(k^2)\frac{g_\sigma^2}{k^2+m_\sigma^2}\bar{h}_1^{(s)}(k^2)+(\frac{g_\rho^2}{k^2+m_\rho^2}+\frac{g_1^2}{k^2+m_\omega^2}\nonumber\\
&&+\frac{g_8^2}{k^2+m_\phi^2})\{h_1^{(\text{v})}(k^2)\bar{h}_1^{(\text{v})}(k^2)(p_1'+q_1')\cdot(p_2'+q_2')+2h_1^{(\text{v})}(k^2)\bar{h}_2^{(\text{v})}(k^2)[(q'_1\cdot q'_2)-(q'_1\cdot p'_2)]\nonumber\\
&&+2h_2^{(\text{v})}(k^2)\bar{h}_1^{(\text{v})}(k^2)[(q'_1\cdot
q'_2)-(p'_1\cdot q'_2)]\}\}c_1\mathcal{F}_{10}(P'\cdot
q',q'^2)\nonumber\\
&&+\int\frac{id^4q'}{(2\pi)^4}(\frac{g_\rho^2}{k^2+m_\rho^2}+\frac{g_1^2}{k^2+m_\omega^2}+\frac{g_8^2}{k^2+m_\phi^2})\{2h_1^{(\text{v})}(k^2)\bar{h}_2^{(\text{v})}(k^2)[q_1'^2q_2'^2-q_1'^2(p'_2\cdot
q'_2)]\nonumber\\
&&+2h_2^{(\text{v})}(k^2)\bar{h}_1^{(\text{v})}(k^2)[q_1'^2q_2'^2-q_2'^2(p'_1\cdot
q'_1)]\}c_2\mathcal{F}_{20}(P'\cdot
q',q'^2)\}\nonumber\\
&&c_2\mathcal{F}_{20}(P'\cdot
p',p'^2)=\nonumber\\
&&\frac{1}{p_1'^2+M_1^2-i\epsilon}\frac{1}{p_2'^2+M_2^2-i\epsilon}\{\int\frac{id^4q'}{(2\pi)^4}\frac{1}{M_1^2p_2'^2}(\frac{g_\rho^2}{k^2+m_\rho^2}+\frac{g_1^2}{k^2+m_\omega^2}+\frac{g_8^2}{k^2+m_\phi^2})\nonumber\\
&&\times\{h_1^{(\text{v})}(k^2)\bar{h}_2^{(\text{v})}(k^2)[(p'_1\cdot
p'_2)(q'_1\cdot q'_2)-(p'_1\cdot q'_2)(q'_1\cdot
p'_2)]+h_2^{(\text{v})}(k^2)\bar{h}_1^{(\text{v})}(k^2)\nonumber\\
&&\times[p_1'\cdot(p_2'+q_2')q'_2\cdot(q'_1-p_1')-(M_1^2+(p'_1\cdot
q'_1))q_2'\cdot(p_2'+q_2')]\}c_1\mathcal{F}_{10}(P'\cdot
q',q'^2)\nonumber\\
&&+\int\frac{id^4q'}{(2\pi)^4}\frac{1}{M_1^2p_2'^2}\{h_1^{(s)}(k^2)\frac{g_\sigma^2}{k^2+m_\sigma^2}\bar{h}_1^{(s)}(k^2)q_2'^2[M_1^2+(p_1'\cdot
q_1')-q_1'^2]+(\frac{g_\rho^2}{k^2+m_\rho^2}+\frac{g_1^2}{k^2+m_\omega^2}\nonumber\\
&&+\frac{g_8^2}{k^2+m_\phi^2})\{h_1^{(\text{v})}(k^2)\bar{h}_1^{(\text{v})}(k^2)(p_1'+q_1')\cdot(p_2'+q_2')q_2'^2[M_1^2+(p_1'\cdot
q_1')-q_1'^2]+h_1^{(\text{v})}(k^2)\bar{h}_2^{(\text{v})}(k^2)\nonumber\\
&&\times[M_1^2(q_1'\cdot q_2')(p_2'\cdot q_2')-M_1^2q_2'^2(p_2'\cdot
q_1')+q_1'^2q_2'^2(p_1'\cdot p_2')-q_1'^2(p'_1\cdot q'_2)(p'_2\cdot
q'_2)]+h_2^{(\text{v})}(k^2)\bar{h}_1^{(\text{v})}(k^2)\nonumber\\
&&\times q_2'^2[p_1'\cdot(p_2'+q_2')q_1'\cdot(q_1'-p_1')
-(M_1^2+(p_1'\cdot
q_1'))q_1'\cdot(p_2'+q_2')]\}\}c_2\mathcal{F}_{20}(P'\cdot
q',q'^2)\},
\end{eqnarray}
where the eigenvalues are different from the eigenvalues in
(\ref{BSE2}) and (\ref{BSE3}). From this equation, we can obtain the
eigenvalues and eigenfunctions which contain the contribution from
the cross terms.

Comparing the terms $p'_{2\lambda}p'_{1\tau}$ in the left and right
sides of Eq. (\ref{BSE2}), we obtain
\begin{eqnarray}\label{BSE4}
\mathcal{F}_1(P'\cdot p',p'^2)
=\frac{1}{p_1'^2+M_1^2-i\epsilon}\frac{1}{p_2'^2+M_2^2-i\epsilon}
\int\frac{id^4q'}{(2\pi)^4}V_{1}(p',q';P')\mathcal{F}_1(P'\cdot
q',q'^2),
\end{eqnarray}
where $V_{1}(p',q';P')$ contains all coefficients of the term
$p'_{2\lambda}p'_{1\tau}$ in the right side of Eq. (\ref{BSE2}). In
this paper, we set $k=(\textbf{k},0)$. Then the fourth components of
momenta of two heavy mesons have no change:
$p'_{10}=q'_{10}=E_1(p_1')=E_1(q_1')$,
$p'_{20}=q'_{20}=E_2(p_2')=E_2(q_2')$. To simplify the potential, we
replace the heavy meson energies $E_1(p_1')=E_1(q_1')\rightarrow
E_1=(M^2-M_2^2+M_1^2)/(2M)$, $E_2(p_2')=E_2(q_2')\rightarrow
E_2=(M^2-M_1^2+M_2^2)/(2M)$. The potential depends on the
three-vector momentum $V(p',q';P')\Rightarrow
V(\textbf{p}',\textbf{q}',M)$. Integrating both sides of Eq.
(\ref{BSE4}) over $p_0'$ and multiplying by
$(M+\omega_1+\omega_2)(M^2-(\omega_1-\omega_2)^2)$, we obtain the
equation of the Schr$\ddot{o}$dinger type
\begin{eqnarray}\label{BSE5}
\left(\frac{b_1^2(M)}{2\mu_R}-\frac{\textbf{p}'^2}{2\mu_R}\right)\Psi_1^{0^+}(\textbf{p}')=\int
\frac{d^3k}{(2\pi)^3}V_{1}^{0^+}(\textbf{p}',\textbf{k})\Psi_1^{0^+}(\textbf{p}',\textbf{k})
\end{eqnarray}
and the potential between $D^{*0}$ and $\bar{D}^{*0}$ up to the
second order of the $p'/M_H$ expansion
\begin{eqnarray}\label{potential1}
V_{1}^{0^+}(\textbf{p}',\textbf{k})&=&
\frac{h_1^{(s)}(k^2)}{2E_1}\frac{g_\sigma^2}{k^2+m_\sigma^2}\frac{\bar{h}_1^{(s)}(k^2)}{2E_2}+h_1^{(\text{v})}(k^2)(\frac{g_\rho^2}{k^2+m_\rho^2}+\frac{g_1^2}{k^2+m_\omega^2}+\frac{g_8^2}{k^2+m_\phi^2})\nonumber\\
&&\times\bar{h}_1^{(\text{v})}(k^2)[-1-\frac{4\textbf{p}'^2+5\textbf{k}^2}{4E_1E_2}],
\end{eqnarray}
where $\Psi_1^{0^+}(\textbf{p}')=\int dp'_0F_1(P'\cdot p',p'^2)$,
$\mu_R=E_1E_2/(E_1+E_2)=[M^4-(M_1^2-M_2^2)^2]/(4M^3)$,
$b^2(M)=[M^2-(M_1+M_2)^2][M^2-(M_1-M_2)^2]/(4M^2)$,
$\omega_1=\sqrt{\textbf{p}'^2+M_1^2}$ and
$\omega_2=\sqrt{\textbf{p}'^2+M_2^2}$. And comparing the terms
$p'_{1\lambda}p'_{1\tau}$ in both sides of Eq. (\ref{BSE3}), we
obtain
\begin{eqnarray}\label{BSE6}
p_2'^2\mathcal{F}_2(P'\cdot p',p'^2)
=\frac{1}{p_1'^2+M_1^2-i\epsilon}\frac{1}{p_2'^2+M_2^2-i\epsilon}
\int\frac{id^4q'}{(2\pi)^4}V_{2}(p',q';P')q_2'^2\mathcal{F}_2(P'\cdot
q',q'^2).
\end{eqnarray}
Setting $\Psi_2^{0^+}(\textbf{p}')=\int
dp'_0p_2'^2\mathcal{F}_2(P'\cdot p',p'^2)$, we obtain the
Schr$\ddot{o}$dinger type equation
\begin{eqnarray}\label{BSE7}
\left(\frac{b_2^2(M)}{2\mu_R}-\frac{\textbf{p}'^2}{2\mu_R}\right)\Psi_2^{0^+}(\textbf{p}')=\int
\frac{d^3k}{(2\pi)^3}V_{2}^{0^+}(\textbf{p}',\textbf{k})\Psi_2^{0^+}(\textbf{p}',\textbf{k})
\end{eqnarray}
and the potential between $D^{*0}$ and $\bar{D}^{*0}$ up to the
second order of the $p'/M_H$ expansion
\begin{eqnarray}\label{potential2}
V_{2}^{0^+}(\textbf{p}',\textbf{k})&=&
\frac{h_1^{(s)}(k^2)}{2E_1}\frac{g_\sigma^2}{k^2+m_\sigma^2}\frac{\bar{h}_1^{(s)}(k^2)}{2E_2}[1-\frac{\textbf{k}^2}{M_1^2}]+h_1^{(\text{v})}(k^2)(\frac{g_\rho^2}{k^2+m_\rho^2}+\frac{g_1^2}{k^2+m_\omega^2}+\frac{g_8^2}{k^2+m_\phi^2})\nonumber\\
&&\times\bar{h}_1^{(\text{v})}(k^2)[-1-\frac{2\textbf{p}'^2+2\textbf{k}^2}{4M_1^2}-\frac{2\textbf{p}'^2+2\textbf{k}^2}{4E_1E_2}].
\end{eqnarray}
In instantaneous approximation the eigenfunctions in Eqs.
(\ref{BSE2}) and (\ref{BSE3}) can be calculated and the eigenvalue
equation (\ref{eigeneq}) becomes
\begin{eqnarray}\label{eigeneq2}
\left(\begin{array}{cc}\frac{b_{10}^2(M)}{2\mu_R}-\lambda&H_{12}\\H_{21}&\frac{b_{20}^2(M)}{2\mu_R}-\lambda\end{array}
\right)\left(\begin{array}{c}c'_1\\c'_2\end{array}\right)=0,
\end{eqnarray}
where the matrix elements
\begin{eqnarray}\label{me1}
H_{12}=H_{21}&=&\int d^3p'\Psi_{10}^{0^+}(\textbf{p}')^*\int\frac{d^3k}{(2\pi)^3}h_1^{(\text{v})}(k^2)(\frac{g_\rho^2}{k^2+m_\rho^2}+\frac{g_1^2}{k^2+m_\omega^2}+\frac{g_8^2}{k^2+m_\phi^2})\nonumber\\
&&\times\bar{h}_1^{(\text{v})}(k^2)\frac{\textbf{k}^2}{E_1E_2}\Psi_{20}^{0^+}(\textbf{p}',\textbf{k}),
\end{eqnarray}
and $b_{10}^2(M)/(2\mu_R)$ and $b_{20}^2(M)/(2\mu_R)$ are the
eigenvalues corresponding to lowest energy in Eqs. (\ref{BSE5}) and
(\ref{BSE7}), respectively; $\Psi_{10}^{0^+}$ and $\Psi_{20}^{0^+}$
are the corresponding eigenfunctions. In Eqs. (\ref{potential1}),
(\ref{potential2}) and (\ref{me1}) the contribution from one-$\pi$
exchange to the potential between two heavy vector mesons has
vanished, but we still give the heavy meson form factors
$h^{(p)}(k^2)$ and $\bar h^{(p)}(k^2)$ in Sec. IV. Then applying the
method as above, we can investigate the alternative $J^P=2^+$
assignment for the \emph{Y} state.

\section{FORM FACTORS OF HEAVY VECTOR MESONS}
To calculate these heavy vector meson form factors $h(k^2)$
describing the heavy meson structure, we have to know the wave
function of heavy vector meson in instantaneous approximation. The
heavy vector meson $D^{*0}$ is regarded as a resonance and its
three-vector wave function in the rest frame has been given
\cite{mypaper3}
\begin{eqnarray}\label{wfV}
\Psi^V(\textbf{p})&=&\frac{2\pi
i}{N^V}\frac{1}{4\omega_c\omega_u}\{exp[\frac{-2\textbf{p}^2-M_1^2-m_c^2+2M_1\omega_c+\Gamma^2/4+(M_1-\omega_c)i\Gamma}{\omega_{D^{*0}}^2}]\nonumber\\
&\times&[\frac{-M_1+\omega_c-\omega_u-i\Gamma/2}{(M_1-\omega_c+\omega_u)^2+\Gamma^2/4}+\frac{M_1-\omega_c-\omega_u+i\Gamma/2}{(M_1-\omega_c-\omega_u)^2+\Gamma^2/4}][\frac{\textbf{p}^2}{3}+\omega_c
M_1-\omega_c^2+\Gamma^2/4\nonumber\\
&+&m_cm_u+i(M_1\Gamma/2-\omega_c\Gamma)]-exp(\frac{-2\textbf{p}^2-m_u^2}{\omega_{D^{*0}}^2})(\frac{\textbf{p}^2}{3}-\omega_u^2+M_1\omega_u+m_cm_u)\nonumber\\
&\times&\frac{M_1-\omega_c-\omega_u+i\Gamma/2}{(M_1-\omega_c-\omega_u)^2+\Gamma^2/4}\}\sqrt{3}\hat\textbf{p},
\end{eqnarray}
where $\textbf{p}$ is the relative momentum between quark and
antiquark in heavy meson, $\hat{\textbf{p}}$ is the unit momentum,
$N^V$ is the normalization, $\omega_{D^{*}}$=1.81Gev
\cite{BSE:Roberts4}, $\Gamma$ is the width of resonance,
$\omega_{c,u}=\sqrt{\textbf{p}^2+m_{c,u}}$ and $m_{c,u}$ are the
constituent quark masses.

\begin{figure}[!htb]
\centering
\includegraphics[scale=1,width=11cm]{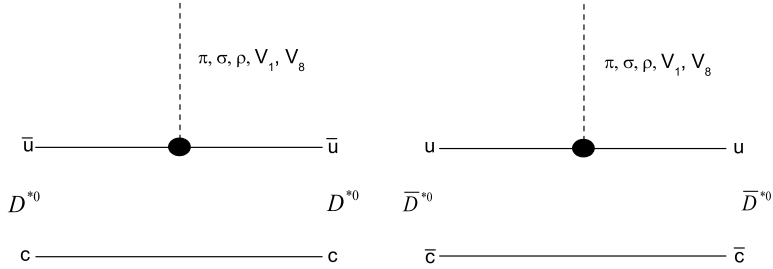}
\caption{\label{Fig2} The vertex function $\Gamma$ in the impulse
approximation. The light meson interaction with $u$-antiquark in
$D^{*0}$ and $u$-quark in $\bar D^{*0}$ is shown.}
\end{figure}

In this paper, we emphatically introduce the form factors
$h^{-}(k^2)$ and $h^{(p)}(k^2)$ derived from one-$\pi$ exchange,
while $h^{(s)}(k^2)$ and $h^{(\text{v})}(k^2)$ have been calculated
\cite{mypaper3}. Firstly, the matrix element of the quark current
$J^-=i\bar{u}\gamma_5 u$ between the heavy meson states (H) has the
form \cite{FF:Faustov1, FF:Faustov2}
\begin{eqnarray} \label{meH}
\langle H(Q)|J^-(0)|H(P) \rangle=\int
\frac{d^{3}qd^{3}p}{(2\pi)^{6}}\bar{\Psi}_{Q}^{H}(\textbf{q})\Gamma(\textbf{p},\textbf{q})\Psi_{P}^{H}(\textbf{p}),
\end{eqnarray}
where $\Gamma(\textbf{p},\textbf{q})$ is the two-particle vertex
function and $\Psi_{Q}^{H}$ is the heavy vector meson wave function
boosted to the moving reference frame with momentum Q. In Fig.
\ref{Fig2} the vertex function $\Gamma(\textbf{p},\textbf{q})$ in
the impulse approximation is shown. The corresponding vertex
function of the meson-quark interaction is given by
\begin{eqnarray} \label{vertex}
\Gamma^{(1)}(\textbf{p},\textbf{q})=\left\{ \begin{array}{cc} \bar{v}_{\bar{u}}(q_{1})i\gamma_5v_{\bar{u}}(p_{1})(2\pi)^{3}\delta(\textbf{q}_{2}-\textbf{p}_{2}) &  $for $D^{*0}$$\\
\bar{u}_{u}(q_{1})i\gamma_5u_{u}(p_{1})(2\pi)^{3}\delta(\textbf{q}_{2}-\textbf{p}_{2})
&  $for $\bar{D}^{*0}$$
\end{array} \right.,
\end{eqnarray}
where $u_u(p)$ and $v_{\bar{u}}(p)$ are the spinors of the quark $u$
and antiquark $\bar{u}$, respectively,
\begin{eqnarray}
u_\lambda(p)=\sqrt{\frac{\epsilon_u(p)+m_u}{2\epsilon_u(p)}} \left( \begin{array}{c} 1 \\
\frac{ {\bf \sigma}\cdot {\bf p}}{ \epsilon_u(p)+m_u }
\end{array} \right) \chi_{\lambda},~v_\lambda(p)=\sqrt{\frac{\epsilon_u(p)+m_u}{2\epsilon_u(p)}}
\left( \begin{array}{c} \frac{ {\bf \sigma}\cdot {\bf p}}{ \epsilon_u(p)+m_u} \\
1
\end{array} \right) \chi_{\lambda},
\end{eqnarray}
with $\epsilon_{u,c}(p)=\sqrt{\textbf{p}^{2}+m_{u,c}^{2}}$   and
\cite{FF:Faustov1, FF:Faustov2}
\begin{eqnarray}
p_{1,2}&=&\epsilon_{u,c}(p)\frac{P}{M_{H}}\pm\sum^{3}_{i=1}n^{(i)}(P)p_{i},\nonumber\\
M_{H}&=&\epsilon_{u}(p)+\epsilon_{c}(p),\nonumber\\
q_{1,2}&=&\epsilon_{u,c}(q)\frac{Q}{M_{H}}\pm\sum^{3}_{i=1}n^{(i)}(Q)q_{i},\nonumber\\
M_{H}&=&\epsilon_{u}(q)+\epsilon_{c}(q),\nonumber
\end{eqnarray}
and $n^{(i)}$ are three four-vectors defined by
\begin{eqnarray}
n^{(i)}(P)&=&\left\{\delta_{ij}+\frac{P_iP_j}{M_{H}[E_{H}(P)+M_{H}]},~i\frac{P_i}{M_{H}} \right\},\nonumber\\
E_{H}(P)&=&\sqrt{\textbf{P}^{2}+M_{H}^{2}}. \nonumber
\end{eqnarray}
In Eq. (\ref{vertex}) the first term  represents the light
pseudoscalar meson interaction with the $u$-antiquark in $D^{*0}$,
while the second term represents its interaction with the $u$-quark
in $\bar{D}^{*0}$. Substituting the vertex function $\Gamma^{(1)}$
given by Eq. (\ref{vertex}) into the matrix element (\ref{meH}) and
then comparing the resulting expressions with the form factor
decompositions (\ref{meofqc1}) and (\ref{meofqc2}), we find
\begin{eqnarray}\label{ffp}
h^{-}(k^2)=\bar h^{-}(k^2)=0.
\end{eqnarray}

\begin{figure}[!htb]
\centering
\includegraphics[scale=1,width=8cm]{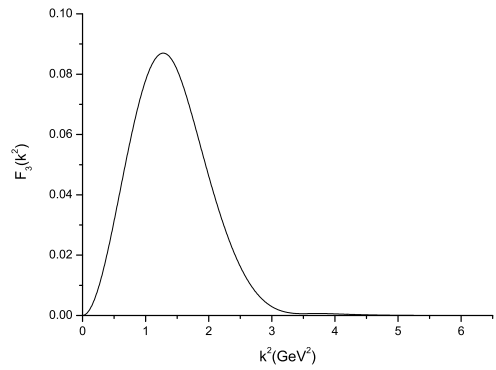}
\caption{\label{Fig3} The form factor for the vertex of heavy vector
meson $D^{*0}$ coupling to $\pi$ meson.}
\end{figure}

Then the quark current becomes $J^+_\mu=i\bar{u}\gamma_\mu\gamma_5
u$ and the matrix element between heavy meson states has the form
\begin{eqnarray}
\langle H(Q)|J_{\mu}^+(0)|H(P) \rangle=\int
\frac{d^{3}qd^{3}p}{(2\pi)^{6}}\bar{\Psi}_{Q}^{H}(\textbf{q})\Gamma_{\mu}(\textbf{p},\textbf{q})\Psi_{P}^{H}(\textbf{p}),
\end{eqnarray}
and the corresponding vertex function is
\begin{eqnarray}
\Gamma_\mu^{(1)}(\textbf{p},\textbf{q})=\left\{ \begin{array}{cc} \bar{v}_{\bar{u}}(q_{1})i\gamma_\mu\gamma_5 v_{\bar{u}}(p_{1})(2\pi)^{3}\delta(\textbf{q}_{2}-\textbf{p}_{2}) &  $for $D^{*0}$$\\
\bar{u}_{u}(q_{1})i\gamma_\mu\gamma_5
u_{u}(p_{1})(2\pi)^{3}\delta(\textbf{q}_{2}-\textbf{p}_{2}) &  $for
$\bar{D}^{*0}$$
\end{array} \right..
\end{eqnarray}
The form factors corresponding to one light pseudoscalar meson
exchange are obtained
\begin{eqnarray}\label{ffv}
&&h^{(p)}(k^{2})=\bar{h}^{(p)}(k^{2})=F_3({\textbf{k}^{2}}),\\
F_3(\textbf{k}^{2})&=&\frac{\sqrt{E_{H}M_{H}}}{E_{H}+M_{H}}\int
\frac{d^{3}p}{(2\pi)^{3}}\bar{\Psi}^{V}(\textbf{p}+\frac{2\epsilon_c(p)}{E_{H}+M_{H}}\textbf{k})\sqrt{\frac{\epsilon_u(p+k)+m_u}{2\epsilon_u(p+k)}}\sqrt{\frac{\epsilon_u(p)+m_u}{2\epsilon_u(p)}}\nonumber\\
&&\times\frac{\textbf{pk}}{(\epsilon_u(p+k)+m_u)(\epsilon_u(p)+m_u)}
\Psi^{V}(\textbf{p}),\nonumber
\end{eqnarray}
where $\Psi^{V}$ is the wave function of heavy vector meson
expressed as Eq. (\ref{wfV}). The function $F_{3}(\textbf{k}^{2})$
is shown in Fig. \ref{Fig3}.

Finally, we obtain the potentials between two heavy vector mesons
for $J^P=0^+$
\begin{eqnarray}\label{potential3}
V_{1}^{0^+}(\textbf{p}',\textbf{k})&=&
-F_1(\textbf{k}^2)\frac{g_\sigma^2}{k^2+m_\sigma^2}F_1(\textbf{k}^2)-F_2(\textbf{k}^2)(\frac{g_\rho^2}{k^2+m_\rho^2}+\frac{g_1^2}{k^2+m_\omega^2}+\frac{g_8^2}{k^2+m_\phi^2})\nonumber\\
&&\times
F_2(\textbf{k}^2)[1+\frac{4\textbf{p}'^2+5\textbf{k}^2}{4E_1E_2}],\nonumber\\
V_{2}^{0^+}(\textbf{p}',\textbf{k})&=&
-F_1(\textbf{k}^2)\frac{g_\sigma^2}{k^2+m_\sigma^2}F_1(\textbf{k}^2)[1-\frac{\textbf{k}^2}{M_1^2}]-F_2(\textbf{k}^2)(\frac{g_\rho^2}{k^2+m_\rho^2}+\frac{g_1^2}{k^2+m_\omega^2}+\frac{g_8^2}{k^2+m_\phi^2})\nonumber\\
&&\times
F_2(\textbf{k}^2)[1+\frac{2\textbf{p}'^2+2\textbf{k}^2}{4M_1^2}+\frac{2\textbf{p}'^2+2\textbf{k}^2}{4E_1E_2}],
\end{eqnarray}
where $F_1(\textbf{k}^2)$ and $F_2(\textbf{k}^2)$ represent the form
factors corresponding to one light scalar and vector meson exchange,
respectively, which have been given in our previous work
\cite{mypaper3}.

The constituent quark masses $m_{c}=1.55$GeV, $m_{u}=0.33$GeV, the
meson masses $m_{\sigma}=0.46$GeV, $m_{\rho}=0.775$GeV,
$m_{\omega}=0.782$GeV, $m_{\phi}=1.019$GeV,
$m_{D^{*0}}=m_{\bar{D}^{*0}}=2.007$GeV and the width of the heavy
vector meson $\Gamma=0.002$GeV \cite{PDG2012}. The equations
(\ref{BSE5}) and (\ref{BSE7}) can be solved numerically with these
potentials in Eq. (\ref{potential3}), and then the eigenvalue
equation (\ref{eigeneq2}) can be solved. Using the same method, we
investigate the molecule state $D^{*0}\bar{D}^{*0}$ with $J^P=2^+$.
Additionally, we investigate the molecule states consisting of two
heavy vector mesons $B^{*0}\bar{B}^{*0}$, which are composed of
bottom quark and $d$-quark. Then the constituent quark masses
$m_{b}=4.88$GeV, $m_{d}=0.33$GeV, the meson masses
$m_{B^{*0}}=m_{\bar{B}^{*0}}=5.325$GeV, the width $\Gamma=0.015$keV
and $\omega_{B^{*0}}=\omega_{\bar{B}^{*0}}=1.81$GeV
\cite{BSE:Roberts4}. Table I shows our results for the masses of
these molecule states.

\begin{table}[ht]
\caption{Masses of charmed and bottom meson molecule states (in
GeV).} \label{table1} \vspace*{-6pt}
\begin{center}
\begin{tabular}{cccccc}\hline
Composite & ~~~ State $J^{P}$ &~~~~ This work
\\ $D^{*0}\bar{D}^{*0}$  & $0^{+}$ & 3.947 &3.970
\\   & $2^{+}$ & 4.142 & 4.147 &~~~4.147 &~~~4.152
\\ \hline $B^{*0}\bar{B}^{*0}$  & $0^{+}$ & 10.459 & 10.476
\\ & $2^+$ & 10.697 &10.699 &~~~10.699 &~~~10.701 \\
\hline
\end{tabular}
\end{center}
\end{table}

For the molecule state $D^{*0}\bar{D}^{*0}$ with $J^P=0^+$, we
obtain two solutions from Eq. (\ref{eigeneq2}). The lower energy
should be the ground state mass of this molecule state which lies
below the threshold and the calculated mass is consistent with the
experimental data, while in the experiment the mass of
\emph{Y}(3940) is $3.943$GeV \cite{SK.CHOI}. This result is
different from Ref. \cite{mypaper3}, this is because in this paper
we consider the SU(3) symmetry. At the same time, the molecule state
$D^{*0}\bar{D}^{*0}$ with $J^P=2^+$ lies above the threshold.
Therefore, we can have a conclusion that if the \emph{Y}(3940) state
is a molecule state composed of $D^{*0}\bar{D}^{*0}$, its quantum
numbers should be $J^P=0^+$. Besides, we can predict that two bottom
mesons $B^{*0}\bar{B}^{*0}$ can be bound by one light meson exchange
as a molecule state with $J^P=0^+$.

\section{Conclusion}
In this work, we assume that the exotic state \emph{Y}(3940) is a
$D^{*0}\bar{D}^{*0}$ molecule state and calculate its mass. The
revised general formalism of the BS wave functions for the bound
states composed of two vector fields is applied to investigate this
issue. Considering one-$\pi$ exchange and SU(3) symmetry breaking,
we construct the interaction potential between $D^{*0}$ and
$\bar{D}^{*0}$ through the heavy vector meson form factors
describing the heavy meson structure. The calculated mass of the
molecule state with $J^P=0^+$ is consistent with the mass of the
\emph{Y} state in experiment, so we conclude that if the
\emph{Y}(3940) state is a molecule state composed of
$D^{*0}\bar{D}^{*0}$, its quantum numbers should be $J^P=0^+$.
Adopting this method, we investigate the molecule states composed of
two bottom mesons.

\bibliographystyle{apsrev}
\bibliography{ref}

\end{document}